# Where's the "Party" in "Multi-Party"?
# Analyzing the Structure of Small-Group Sociable Talk

Paul M. Aoki,[2] Margaret H. Szymanski,[1] Luke Plurkowski,[1] James D. Thornton,[1]
Allison Woodruff [2] and Weilie Yi [3]

| [1] Palo Alto Research Center | [2] Intel Research Berkeley | [3] University of Rochester |
|---|---|---|
| 3333 Coyote Hill Road | 2150 Shattuck Ave., Ste. 1300 | Dept. of Computer Science |
| Palo Alto, CA 94304-1314 USA | Berkeley, CA 94704-1347 USA | Rochester, NY 14627-0226 USA |

## ABSTRACT

Spontaneous multi-party interaction – conversation among groups of three or more participants – is part of daily life. While automated modeling of such interactions has received increased attention in ubiquitous computing research, there is little applied research on the organization of this highly dynamic and spontaneous sociable interaction within small groups. We report here on an applied conversation analytic study of small-group sociable talk, emphasizing structural and temporal aspects that can inform computational models. In particular, we examine the mechanics of multiple simultaneous conversational floors – how participants initiate a new floor amidst an on-going floor, and how they subsequently show their affiliation with one floor over another. We also discuss the implications of these findings for the design of "smart" multi-party applications.

## Categories and Subject Descriptors

H.5.3 [**Information Interfaces and Presentation**]: Group and Organization Interfaces – *theory and models.*

## General Terms

Design, Human Factors

## Keywords

Sociable talk, conversation analysis, small group interaction.

## 1. INTRODUCTION

Spontaneous *multi-party interaction* – conversation among groups of three or more participants – occurs within countless everyday sociable encounters: family dinners, hallway chats, parties, friendly get-togethers. However, studies of social interaction rarely have a primary focus on the organization of multi-party interaction. Pragmatic factors (such as a lack of speaker-separated recordings, prevalence of speech overlap and disfluencies, and frequent shifts of topic or reference) reduce the intelligibility of such interactions, making both qualitative [10] and quantitative

[27] analysis very difficult. Unsurprisingly, most work examines the more easily-studied case of two-party (dyadic) conversation.

Nevertheless, researchers in ubiquitous computing environments are gradually recognizing the need to better understand multi-party interaction. Automated analysis of multi-party interaction has recently been applied in environments such as co-present meetings [17,27,30], co-present informal interactions [4,8,20], and remote interaction via audio or video links [2,21,32]. This nascent line of research seeks to enable applications such as:

- tracking conversational activity and meeting status in "smart meeting rooms" for archival and assistive purposes [17,27,30];
- discovering long-term social interaction patterns to inform social network analysis ("social dynamics") [4,8,20];
- automating volume control of remote conferencing systems to facilitate spontaneous talk [2,32], or
- resolving addressing ambiguities in spoken human-robot interaction [5].

What is still largely missing, however, is a detailed understanding of the structural characteristics of sociable, multi-party talk that tend to distinguish it from dyadic talk and workplace meetings. As social scientists have widely observed [9-11,23,24,28,31], sociable interactions frequently result in multiple, simultaneous conversations. In most of these applications, then, a key concern (albeit one usually left for future work) is the development of machine learning models to recognize "who is talking to whom" – that is, which participant is party to which conversation.

This concern with multiple simultaneous conversations is central to our research agenda. Inspired by early work on audio-only media spaces [1], we set out to design mobile audio communication services for small social groups, particularly in the teen and young adult markets. Based on our design fieldwork of mobile lightweight audio communication in this demographic [35], we previously designed and implemented a multi-party audio space system that facilitated multiple, simultaneous conversations between mobile users [2]. The goal is to enable highly spontaneous, sociable interaction – the kind that is so common and enjoyable in face-to-face interaction but stifled by standard monaural audio technology – for mobile users. The system works by (1) applying machine learning models to identify the participants of the conversation(s) and (2) increasing intelligibility within each conversation by providing a customized audio mix to each participant. Because of the mobile use scenario, the system must work using only audio communication data and a minimal user interface. Informed by studies of an initial prototype, we have recently returned to this work with the goal of improving the models and increasing the prototype's recognition accuracy.

The paper has two main areas of contribution. First, from a social science perspective, we report on a qualitative study that applies the methods of conversation analysis [22,23] to examine the structure of small-group sociable talk. Detailed analysis of recordings of such interactions enables us to describe some important ways in which participation is organized in spontaneous, multi-party interaction. (In that sense, it resembles qualitative CSCW research that examines how people manage participation in multiple physical/virtual environments, multiple conversations, etc. [7,16,31].) Since our ultimate goal is to facilitate automated analysis, these descriptions emphasize structural and temporal elements – what we will later call *participation sequences* – as opposed to semantic elements. Second, from a technological perspective, we discuss the implications of our social science findings for the design of "smart" multi-party applications such as those described above. In particular, we describe how the qualitative phenomena can be incorporated into models of multi-party interaction as well as the user interfaces of applications that include such models.

The paper is organized as follows. In the next section, we provide some background on multi-party interaction. We then turn to sections describing our study and its findings, respectively. A discussion of implications for design and methodology precedes our summary and conclusions.

## 2. MULTI-PARTY INTERACTION

Social scientists have long remarked on the dynamic, fluid nature of sociable conversation, as when Simmel characterized it through "the lively exchange of speech" ([29], p.52); our goal has been to move past glosses and understand how this "lively" fluidity is actually achieved by participants. While our approach draws on conversation analysis [22,23], we have found it useful to think about alternative approaches to the question of "participation" in interaction. We (briefly) review some of these in turn, ultimately developing the notion of participation sequence that we will use to frame our analysis in Section 4.

One class of approaches centers on individual participants. The best-known of these is the *participation framework* [14]: a categorization of each person's participation status in relation to a given social event (such as a particular utterance), including specifications as to their appropriate conduct given this status. Such typologies of "role or function" ("speaker," "addressed recipient," "bystander," etc.) are generally helpful in describing participants' immediate actions. Being participant-centric, they do not shed much light on exactly "how" roles change over the course of an interaction.

Another class of approaches centers on collections of participants, grouping people by "who is talking to whom" but defining it in different ways. A collection-centric approach defined in this way has a stronger temporal aspect than a participant-centric approach, since grouping people in this way necessarily considers multiple utterances. In many cases, particularly in analyses of workplace meetings, multi-party interaction is conceived as a stable central activity, even when multiple people are speaking at once (e.g., [9,10,27]). Where it is explicitly recognized that multiple, simultaneous conversations do occur, they are often dichotomized as "main" and "side" conversations, with one "dominating" and the other "subordinate" in terms of duration and volume (e.g., [14]). Surprisingly few studies (e.g., [28]) explicitly recognize that stable simultaneous conversations occur at all. A useful abstraction of this kind is Shultz *et al.*'s notion of *primary* and *secondary* participation [28], which recognizes that sub-groups of (primary) participants often do the bulk of the speaking for extended periods.

A third approach that we have found more useful is to examine multi-party data in terms of *turn-taking systems* [23] – to view participation as "a temporally unfolding, interactively sustained, embodied course of activity" [15] – instead of categorizations. This approach assumes a *social action*, overwhelmingly a turn at talk, as the basic unit of analysis. How activities are produced from these units is described below.

One of the fundamental structures of conversation is the organization of taking turns at talk. The turn-taking system specifies an organization to the ways in which transition occurs from current speaker to next speaker. Overwhelmingly, the next speaker positions his or her turn at a place where the current speaker has produced an understandably complete utterance or turn-constructional unit (TCU); a single turn at talk may consist of several TCUs. Recipients are not passive listeners but incipient speakers, continuously monitoring current talk to project the completion of the current speaker's TCU or a transition-relevance place (TRP) where speaker change may occur.

One can refer to a given social encounter as having one or more *floors*, where each floor is instantiated by the engagement of the turn-taking system. Conversation analysts, often stereotyped as conceptualizing turn-taking through a strict model of one-at-a-time speech [9,10,27], have in fact consistently argued that such *schismings* [11,23,24] into multiple floors occur (especially amidst spontaneous, non-task-oriented conversation). In this paper, we will say that speakers' actions produced within a particular turn-taking system *affiliate* them with that floor. Whereas a single floor is visible by the operation of a single turn-taking system, a successful schisming is visible by the sustained operation of an additional turn-taking system in parallel.

All three approaches come together in the following notion. In our analysis, we will aim to locate speakers according to floor by describing what we will call a *participation sequence* – the actions that they produce to empirically demonstrate an orientation toward another speaker to establish a specific floor. A speaker is member to a particular participation sequence as long as their turns at talk are related to adjacent turns at talk in the sequence. Affiliation with a floor ceases when an individual ceases producing turns at talk that orient to it. As the participation sequence unfolds, participants display their affiliation or disaffiliation by the presence or absence of their actions over time.

Multiple floors are a notable feature of spontaneous, sociable interaction; enabling "more sociable" and "party-like" interactions will involve understanding how multiple floors are organized. Specifically, we examine how participants become party to a conversational floor through schismings and affiliation.

## 3. METHODS

To investigate phenomena relating to multi-party sociable interaction, we produced and analyzed a corpus of audio

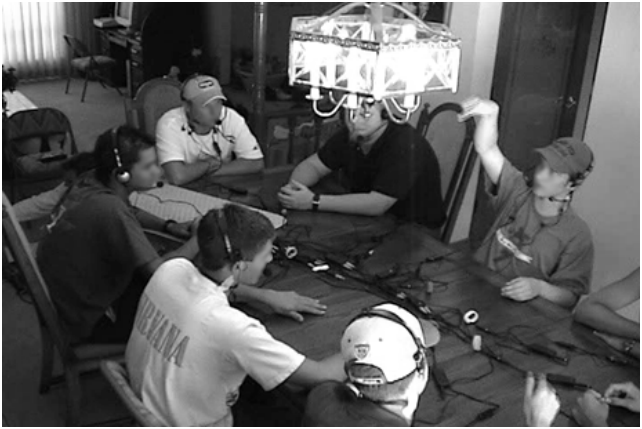

**Figure 1. Data collection from co-present social interaction.**

recordings.[1] In this section, we describe our data collection and analytic methods, respectively.

It is important to remember that this work does not follow an experimental paradigm. The core of what we report in this paper is a qualitative analysis of spontaneous, naturally-occurring human interaction. The care taken in the technical aspects of data collection is meant to facilitate its use as machine learning training data, not the collection of reportable descriptive statistics, for which we do not have a sufficiently large or diverse corpus.

## 3.1 Data Collection

The findings of Section 4 are based on a corpus of multi-party, co-present, sociable interaction, recorded in multiple sessions. We recorded both audio (using multiple sensors [33]) and video for each session, each of which consisted of an interaction, one hour in length, among a group consisting of 8-10 participants. Each session's group had a different composition, though some participants were present in more than one session.

We recorded all sessions in closed configurations in a quiet sitting area − e.g., a circular arrangement of couches in a family living room, or an arrangement of seats around a table in a family dining room (Figure 1). This had two implications. First, the setting itself provided relatively little in terms of local resources [22] for talk, such as notable events or physical features. (Further, the protocol did not involve resources such as prompting or tasks.) Second, while participants were not strictly equidistant to each other, all participants could potentially hear each other and have access to each other's non-verbal communication (gesture, posture, gaze, etc.).

As we have expanded our corpus, our goal has been to make it increasingly reflective of sociable interaction in our target (teen and young adult) demographic. We initially collected two *pilot* sessions from co-workers in our research organization. Session groups were mixed-gender. Participants were U.S. residents (all but two were native English speakers), had relationships ranging from close collaborators to acquaintances, were of ages from 20 to 40, and were uncompensated. We have since collected seven *youth* sessions from members of the second-degree social network (i.e., up to "friend of a friend") of a single teenager. Session groups were both single- and mixed-gender. Participants were U.S. residents (all were native English speakers), had relationships ranging from romantic couples to acquaintances, were of ages from 14 to 24, and were compensated. Two pilot and two youth sessions (a total of 30 unique participants) have been fully analyzed as described below.

## 3.2 Analytic Methods

We now turn to the analytic methods on which the findings of Section 4 are based. Our qualitative analysis draws primarily on the research findings and methods of *conversation analysis*, an inductive process for analyzing how human interaction is organized into sequences of action or systematic practices [22,23]. To identify these systematic practices, sections of audio data (and video data, where appropriate) are transcribed in detail and then analyzed, first individually then comparatively, to produce collections of qualitatively similar phenomena. That is, the analyst first examines how sequences of action are organized and situated in particular instances of activity ("individually"), then abstracts features that generalize across various instances ("comparatively"). Collections of phenomena produced by this process are inductive and data-driven; while motivated by specific research questions, they encompass phenomena actually observed in a corpus (as opposed to approaching the data with taxonomies posited by theory *a priori*). A canonical example is the description by Schegloff & Sacks of a collection of systematic practices that become relevant as participants bring telephone conversations to a close [26].

As part of this qualitative analysis process, we manually produce a set of floor *labels*, i.e., a partitioning of each participant's activity according to floor affiliation. Our technical reason for this is to produce labeled training data for supervised machine learning, i.e., to enable extraction of statistical features of turn organization relating to a single floor, between floors, and across floors [2]. Of relevance to this paper, however, is that explicitly tracing floor affiliations enables us to locate and describe the systematic practices that become visible as new floors emerge and participants affiliate with different floors.

Identifying how interactions are organized into sequences of actions is a key element of conversation analysis, and floor labeling can be seen as a product of the exhaustive execution of this process. We decide that a new floor arises when a participant responds verbally to another participant's conversation-initiating turn, engaging the new turn-taking system. A heuristic decision that this is a new, parallel turn-taking system can be made from the fact that such a system − starting with this initial two-turn sequence − operates independently of the previously-established turn-taking system and results in sustained simultaneous speech (overlap) with it. (We expand on this in Section 4.2.) However, conversation analytic evidence of the independent and parallel operation of turn-taking systems accumulates quickly to confirm (or disconfirm) this decision. For (off-line) labeling purposes, we timestamp the emergence of the new floor at the onset of the initial participant's turn; timestamp the first responsive participant's affiliation with the floor at the onset of their first turn; and so on. Similarly, as participants visibly orient to other floors, the completion of their final turn marks the end of their previous affiliation. (The completion of a participant's last utterance does not necessarily coincide with their internal shift in attention away from that conversation, but attention shift may not be visible to an analyst.)

---

[1] Previous speech corpora were unsuitable for our purposes. Quantitative analysis of turn-taking, even in larger social groups, is generally of dyadic interactions (e.g., [4,6,8,18,34]). Qualitative studies (e.g., [11]) collect data in a manner unsuitable for our eventual goal of automated analysis. The exceptions are the recent multi-party meeting corpora collected by speech recognition researchers (e.g., [17,27,30]), but these are all of workplace meetings rather than of sociable interaction.

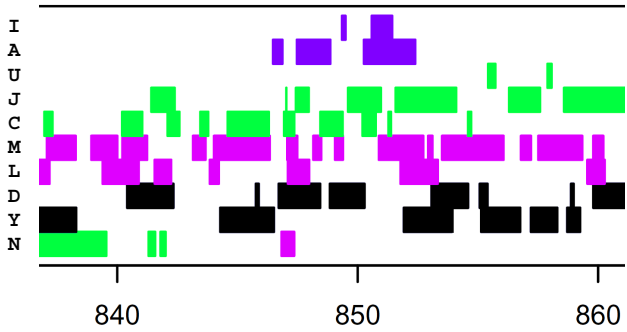

**Figure 2. Voice activity of ten participants with four floors.**

The qualitative analysis is supplemented by visualizations of the individual participants' speech segments as measured by an energy-based *voice activity detector* (VAD) (Figure 2). (VADs are commonly used in speech systems and analyses – see, e.g., [6,18,27,34].) Breaks between VAD segments visually identify silent/quiet periods, which is useful since a TRP is often marked by a beat of silence in conversation. Similarly, VAD segments can also be used to identify periods of simultaneous speech. Figure 2 depicts the VAD segments of ten participants (vertical axis) over time (horizontal axis, in seconds); segments of the same color/shading indicate affiliation with the same floor. As we will see in Sections 4 and 5.1, VAD diagrams are useful in refining the qualitative analysis.

Before we move on to the findings of our qualitative analysis, we note some general properties of our corpus that were revealed through the manual floor labeling. First, multiple-floor configurations are not rare. Figure 2 illustrates this, with up to four floors simultaneously active. In every labeled session, the number of simultaneously active floors ranges from one to a maximum of three or four (with a time-weighted mean of 1.79 for all sessions). Second, floors are very dynamic. The labeled pilot sessions contain 10 and 19 distinct floors per hour-long session. The labeled youth sessions contain 52 and 70 distinct floors per hour-long session. While every session has floors lasting under 10 seconds and over 15 minutes, floors in the pilot sessions tend to last somewhat longer than those in the youth sessions (median of 91 and 44 seconds, respectively). While this shows (predictably) that different groups show diversity of spontaneity, it also highlights the fact that the entire corpus involves many, relatively short-lived configurations of participants, reinforcing observations about the fluidity of sociable talk.

## 4. ANALYSIS

In this section, we discuss a number of qualitative phenomena related to the organization of conversational floors in multi-party talk. The first four phenomena address ways in which a new floor is initiated in our multi-party data, significantly extending the conversation analysis literature on this topic – in particular, three of the four have not been previously analyzed in this way. The last two phenomena describe how speakers demonstrate affiliative action once a floor has been established. (We do not discuss how speakers end affiliation with a floor, as this generally becomes visible through some combination of (1) lapse in participation and (2) affiliation with a different floor. Hence, initiation and affiliation are clearly the most important organizational aspects for conversational modeling.)

The phenomena discussed in the remainder of this section examine speakers' actions within participation sequences in the corpus described in Section 3.1. Each is labeled "(P*n*)" to simplify references to them in subsequent sections. Phenomena are illustrated by transcript excerpts where it is necessary. The transcription conventions are based on Jefferson's ([3], pp.ix–xvi) and are outlined in Table 3. We have inset VAD diagrams like that shown in Figure 2 into most of the excerpts, which have been placed together on one page so these diagrams can be compared visually (relevant at the end of Section 4.1).

### 4.1 Schisming

Here, we examine how speakers' actions can provide resources for schismings, the emergence of an additional floor amidst an on-going floor(s), to occur in multi-party interaction. The first phenomenon discusses how schisming that occurs in a single floor causes two floors to emerge; in this case a speaker initiates schisming in an explicit, targeted manner. Although this phenomenon has been extensively discussed elsewhere [11], it provides an important point of contrast for the three newly-analyzed schisming phenomena that often occurred amidst multiple simultaneous floors. In our data, schisming may not feature a targeted initiation – indeed, the first action may not have been intended to cause a schisming at all.

### (P1) Schisming by Schism-Inducing Turn

When we think of the schisming of one conversational floor into two, we often think of those cases in which the schisming clearly arises from one participant's actions designed to establish the recipiency ("get the attention") of one or more others. Such actions are designed to stand apart from a currently-engaged turn-taking system and, if taken up, result in the engagement of new one. Egbert [11] describes the *schism-inducing turn* (SIT) as having three characteristics: (1) it causes a change in topic, (2) it is a first pair part action that initiates a new sequence and implicates a responsive second pair part action (ex. question – answer) [26], and (3) it directly targets a specific recipient or recipients. Here, *target* means that the design of a SIT draws on the usual resources available to speakers to maximize the chance of a successful bid for recipiency: positioning of the SIT "in the clear" (i.e., without overlap) in a TRP, the use of direct verbal address (e.g., names and pronouns), employment of gaze and gesture in copresence, etc.

Excerpt 1 is a canonical example of a SIT and its uptake. As the excerpt begins, two other dyadic conversations are in progress (not shown), and the featured participants are orienting to one of these other conversations. In line 1, participant N addresses participant C (who has not spoken for quite some time) with a SIT – repeatedly calling C by name (targeting), then asking him a question (first pair part) on a new topic (topic shift). Although the SIT in line 1 is targeted at C, J elaborates on N's initial turn by soliciting confirmation of a candidate answer to N's question in line 5. After C and N clarify the time frame of N's initial question in lines 2 and 3, C rejects J's candidate answer ("nohh") and responds to N's question ("I had a garage sale") in line 4. The new floor created by N, C and J continues in lines 8-11.

While SITs such as that shown in Excerpt 1 are common and familiar occurrences that have been well-described in the literature, across our corpus, only 27 of the 153 floors identified were initiated with SITs. Overwhelmingly, floors were initiated by turns produced *within* the engaged turn-taking system, resulting in the engagement of another turn-taking system. Here, we discuss three such types of turns that, when produced in certain participatory contexts, can occasion a schisming amidst multiple simultaneous conversations.

*(P2) Schisming by Toss-Out*

It is often observed that participants in sociable talk frequently produce turns in a "tossed out" manner [9,10,28] that – somehow – does not "require response or acknowledgement" [28]. Here, we go past this gloss and say that what qualifies them as "tossed out" is that they are, by turn design, a type of action that does not do the social work of requiring response or acknowledgement; such turns, which commonly take the form of announcements, noticings or outlouds, are known to have less strength in soliciting a response as a first pair part than (e.g.) a question.

In our corpus, the *toss-out* turn type often results in schisming. This is particularly interesting because the features of a toss-out are almost diametrically opposed to a canonical SIT, and instead are produced like nearly any other contribution to the ongoing conversation: (1) it is topic-relevant to the in-progress conversation, (2) it is organizationally responsive to the in-progress conversation, occurring within its turn-taking system, and (3) it does not directly target a specific recipient or recipients. Like SITs, toss-outs are first pair parts and are often initiated at a TRP, but it should be remembered that turns responsive to an in-progress conversation are overwhelmingly initiated at TRPs.

A toss-out can result in three different outcomes. First, no one may respond to the toss-out at all. Second, someone may respond to the toss-out and follow its trajectory within the current conversation. In both of these cases, no new floor emerges. In the third outcome, however, someone may respond to the toss-out creating a new turn trajectory in parallel to the conversation in which the toss-out was produced; this results in the emergence of a new conversational floor.

Excerpt 2 illustrates how a single floor becomes two distinct floors as the result of a toss-out. T describes a friend's car refurbishment project (a "'68 Nova") in line 1, receiving positive assessments and a follow-up question from H, Z and S in lines 3-8. N then announces, "I need to sell my Mustang" in line 9 (marked by arrows in the transcript and diagram) with all the features of a toss-out: topically coherent with the on-going conversation, produced at a TRP, and untargeted at any particular recipient. Although N positions his announcement within a TRP, H displays his orientation to it as a competing turn within the turn-taking system in operation as evidenced by his cut off and restart in line 10 and subsequent silence in line 11. At this interactional crossroads, the participants diverge: J responds to N's toss-out in line 12 to engage a new floor, and T and H continue the "'68 Nova" conversation.

In Excerpt 1, a toss-out turn similar to the one in Excerpt 2 results in a schisming. Recall how in lines 1-5, C, J and N are discussing sleep and morning activities. In line 6, A produces a telling about his own morning activity – his failure to unset snooze mode on his alarm clock. A's telling is tossed into the conversation at an opportune moment: the TRP at the boundary of a sequence (N's question in line 1 – C's answer in line 4). A begins his TCU in the clear, increasing recipients' ability to hear and respond to it. In line 7, M responds to A's telling as talk about the garage sale continues in the other floor. It should also be noted that A's toss-out turn results in the emergence of a fourth floor, as two floors were already existent before N's successful SIT.

Excerpt 1 clearly illustrates the difference between a canonical SIT and schisming by toss-out. N's SIT in line 1 is distinctly marked as the first action in a conversation – it effectively gains recipiency and implicates the response of the target recipient through the production of the strongest first pair part action: a question. A's turn in line 6 produces different constraints on its recipients given its action (telling) and position amidst an already in-progress interaction.

Comparing their interactional features, A's toss-out is a weaker bid for response than N's SIT. Together, the lack of a targeted recipient and its weaker status of a toss-out's action reduce the relevance of a responsive action. Socially, the absence of a response to a toss out is much less noticeable than an unanswered SIT. Further, whereas speakers use SITs specifically to initiate a new conversation, toss-out producers are relatively agnostic with regard to the continuity in the current conversation or the emergence of a new one. Instead, it is the toss-out turn responders – who coincidentally are often secondary participants in the sense described in Section 2 – who define the toss-out's trajectory as an initiating action in a new conversational floor.

*(P3) Schisming by Aside*

Schisming can also occur in sociable talk through the use of an "aside." We now characterize the *aside* turn type in conversation analytic terms, as we did for toss-outs; in fact, asides are similar to toss-outs, but with several key differences.

Asides, like toss-outs, are topic-relevant for the current conversation and do not strongly implicate a response. Asides are produced in a marginal way to the on-going conversation: (1) they are often positioned in overlap with the on-going conversation (i.e., not at a TRP), and (2) they are produced in a subdued voice. As we will see, this "marginality" can be a resource in schisming.

Excerpt 3 illustrates how schisming occurs through the use of an aside. As the excerpt begins, the participants are discussing how lifeguards at the local swimming pools and reservoirs make the racist assumption that people of visibly African ("Blacks") and Hispanic ("Mexicans") descent do not know how to swim. In line 6 (marked by the arrow), S produces a racist joke in the form of an aside that implicitly references the stereotype that "Mexicans" cross into the U.S. by swimming across the Rio Grande. In line 10 and the beginning of line 12, S continues to think aloud about the scenario. J's responsive laughter in line 13 converts the aside into a successful schisming, interactionally visible by S's marked increase in volume and prosodic stress on the (already understood) punch line of his joke: "swam over here."

By design, asides are produced to be marginal to the in-progress conversation, consequently they are inclined to be taken up in a new conversation. Because asides are audibly differentiated from the on-going conversation, speakers can use it as a resource to recipient design their talk for secondary participants (who are mainly listeners). That is, producing the turn in a soft voice (especially when produced in overlap) targets the action towards people who are not attending to the main conversation as primary participants. As with toss-outs, speakers responding to asides transform it into the first turn of a new conversation. With asides the interactional work of gaining recipiency is shifted from the initiator to the responder (e.g., not "Listen to me" but rather "I heard you"). Further, the aside turn type provides a way to participate in an on-going conversation without interrupting a sequence of action in progress.

## (P4) Schisming by Retro-Sequence

Thus far, we have seen how three types of first pair part actions – SITs, toss-outs and asides – can result in a new conversation. A fourth type of turn, one that is not designed or produced as a first pair part initiating action, can also result in a new conversation. We are not characterizing this turn type for the first time as we did toss-outs, but its role in schisming has not been described.

In general, *retro-sequences* [25] are turn sequences in which a first turn "goes by" with an initial status until a subsequent turn implicates a different status.

> [T]he source engendered nothing observable – indeed was not recognizable as a 'source' – until the later utterance/action [from a different speaker], billing itself as an 'outcome', retroactively marks it as such. Their 'firstness' follows their outcome, though their occurrence preceded it. ([25], p.235)

A simple example is when a comment not produced as a joke is transformed into humor by its recipients' laughter.

In our corpus, schisming by retro-sequence follows from a "source" turn such as an observation or a joke. The first turn of a retro-sequence schisming is actually a second pair part action, a response to a prior turn in the current conversation. These source turns contribute to the continuity of the in-progress conversation in which they occur and are not designed to be taken up in a new conversation. They are available, however, for participants to use as launching grounds for a new conversation. Retro-sequence schisming occurs when speakers respond to a prior responsive turn (e.g., an answer) with a topic-relevant utterance not produced within the turn-taking constraints of the current conversation.

As an example of retro-sequence schisming, consider Excerpt 4, in which H has just told a story about a fistfight at the swimming pool where he works as a lifeguard. N pursues more detail about whether the lifeguards got involved in the fight in line 1, prompting H to extend his telling in lines 3-4. The participants provide varied responses: choral laughter [19] (lines 5-11), an assessment (line 12) and an alternate formulation of H's telling (lines 13). As H continues in line 14, Z takes up N's alternative punch line in line 15, responding in overlap with H. (The retro-sequence parts are marked by arrows.) Note that N's punch line ("Get my gu:n?") was not directed at Z and did not directly implicate a response; its status as the "source" of Z's subsequent uptake is produced retroactively by Z's uptake. (Contrast this with W's very similar punch line, delivered at Line 16, which is not taken up by anyone and hence is never converted into a "source" of a subsequent schisming.)

Unlike SITs, toss-outs and asides, retro-sequence schisming occurs when a responsive turn produced in an on-going conversation is transformed by a subsequent responsive action into the initiating turn of a new conversation.

## Schisming: Discussion

To summarize (P1)-(P4) above, each describes a sequence type that results in schisming, but the archetypical first turn of each is:

**Table 1. Schisming first turns.**

|  | a… | topic… | targeted… | positioned… |
|---|---|---|---|---|
| (P1) SIT |  | -changing | directly | in the clear |
| (P2) toss-out | FPP | -relevant | indirectly | in the clear |
| (P3) aside |  |  |  | in overlap – subdued delivery |
| (P4) retro-sequence | SPP |  | – | in the clear ("outcome" turn in overlap) |

where FPP/SPP indicates first pair part and second pair part.

For the purposes of our later discussion of design implications, the most relevant of the distinctions in Table 1 will be targeting (SIT vs. the others) and volume/positioning (aside vs. toss-out and retro-sequence). The first two (pair part type and topic relevance) are semantic distinctions, whereas positioning is directly visible in a participation sequence (as suggested by the arrows in each of the inset VAD diagrams) and local utterance properties such as volume and targeting can potentially be detected without using full natural language processing (e.g., using speech energy measurement and wordspotting).

## 4.2 Affiliating

The practices described above do not cause floors to emerge instantaneously, fully-formed and unchangeable. A floor is defined by its constituent sequences of action, and participants' affiliation changes over time; participants may begin to orient to new floors after they emerge, and cease to orient to ongoing floors. In co-present communication, much of this is accomplished through non-verbal means such as gaze and posture. In mediated communication, more tends to be done through verbal means. We describe two such means here.

## (P5) Affiliating by Turn-Taking

We can infer participation in floors from verbal communication, using speech to make inferences about the organization of turn-taking. For example, when we can see the operation of a new turn-taking system, we have evidence of a schisming. Moreover, multiple, simultaneous conversations can be distinguished from one other by comparing the organization of turn-taking in each conversation [2]. In a single conversation, sustained simultaneous speech is relatively infrequent; when overlap does occur, one or more of the overlapping speakers typically drop out [23,24]. When two simultaneous conversational floors are on-going, participants orient their participation to the turn-taking organization of their own floor, not to the other floor. This has two important implications: (1) in one floor, participants' turns are produced with minimal gap or overlap; that is, TCUs are initiated in the TRPs of their own conversation, and (2) when two or more floors exist, participants' talk will overlap much more than if they were all participating in a single floor.

In quantitative terms, we can make inferences about an individual's floor participation by looking for patterns of (1) alignment of turns within a hypothesized floor and (2) sustained overlapping talk with all other hypothesized floors. (If a participant produces no or very few utterances, as in the case of secondary participation, there will be little information.) The exact distribution of "typical" alignment and overlap, both within and between floors, can be derived from theoretical models [6,18,34] or from empirical (data-driven) models. Variations of this basic idea have been applied in several of the systems mentioned in the introduction [2,4,8,20] – even in the absence of an understanding of (P1)-(P4), turn-taking alone has been shown to be very useful in modeling floor organization.

### (P6) Affiliating by Coordinated Action

While the basic organization of turn-taking is perhaps the most visible and stable signature of a floor, it is not the only one. Participants show their affiliation with a floor through means other than turn-taking (i.e., overlap and alignment). In fact, some of these other means run counter to those of turn-taking.

In our corpus, participants frequently demonstrate affiliation with a given floor through finely-coordinated actions, often short vocalizations. However, this coordination is not aimed at minimizing gaps and overlaps. Instead, it often results in the maximal overlap of these short utterances. We see this clearly in Excerpt 4, where all eight participants engage in some kind of response to the punch line of H's story in lines 5-13; most engage in highly-coordinated simultaneous laughter. (The inset VAD diagram in Excerpt 4 represents the same time period shown in the transcript. Because people laugh differently and with different loudness, the two do not correspond perfectly – but the overall structure can be seen, particularly the onset of laughter.)

This example reflects the fact that one of the ways participants tease apart the confusion caused by overlapping talk in a floor is to analyze its characteristic features. Each form of overlapping action – shared laughter following a joke [13], choral and cooperative turn completion [19], as well as others described in the conversation analytic literature – is a coordinated achievement representing a high degree of affiliation among its participants. Consequently, these specific types of overlapping actions "override" the "rule" of minimizing gap and overlap.

Unlike affiliation by turn-taking, affiliation by coordinated action can apply even in cases of secondary participation. For example, during the period immediately preceding the punch line in Excerpt 4, A, S, T and W do not contribute a turn at talk to the conversation, but they all participate through shared laughter.

### Affiliating: Discussion

Affiliation-related phenomena are different from the schisming-related phenomena in that they do not describe specific sequence types. We might summarize (P5)-(P6) in a different way:

**Table 2. Floor-affiliating actions.**

|  | overlap is… | delivery is… | participation by… |
|---|---|---|---|
| (P5) turn-taking | "minimized" | normal | (mainly) primary |
| (P6) coordinated action | sustained | emphasized | primary/secondary |

As with positioning in Table 1, we note that degree of overlap is directly visible in a participation sequence. Similarly, like subdued delivery in Table 1, emphasized delivery can potentially be detected without natural language processing.

## 5. IMPLICATIONS

We now draw on our analysis to inform the design and development of "smart" multi-party applications, such as those described in the introduction, that model participation in multi-party interaction. We have divided these implications into three main categories: the process of producing training data, specific user interface features, and general observations on modeling of multi-party interaction. Where appropriate, points are cross-referenced with Section 4 phenomena using the notation "(P*n*)".

### 5.1 Tools for Qualitative Analysis

A system attempting to model conversational behavior using supervised machine learning techniques (our own [2], or others mentioned in the introduction, e.g., [4,5,8,20,21,27,30]) requires collection and labeling of training data. Making the labeling task more efficient will speed development time and reduce costs.

We found simple VAD diagrams, such as that seen in Figure 2, to be a very useful aid in the qualitative analysis relating to floors. Certain phenomena are visually salient that might not otherwise be noticed with the (more faithful but complex) audio waveform view provided by most tools or a textual transcript. Compared with looking at very long stretches of multi-track audio data, VAD diagrams provide a quick way to (1) identify candidate phenomena and locate additional instances and (2) locate recording anomalies and possible labeling errors. For example, the operation of a turn-taking system (P5) is often visible in the VAD diagram. In Figure 2, participants J, M and Y frequently overlap with each other for extended periods; as a result, they seem likely to be (and, in fact, are) participants in different floors. Knowing this alone is helpful in making an initial guess as to which other participant(s) a given comment is responsive. Similarly, patterns of coordinated action (P6) can be seen, such as the choral laughter of Excerpt 4. While the patterns are not always unambiguous or exactly what a transcriber would produce, the examples should make the utility of the method clear.

### 5.2 Design and User Interface

The phenomena described in the previous section (and somewhat reductively summarized in Table 1 and Table 2) suggest a number of user interface features for applications that model and adapt to conversational floor participation. We describe two such classes of system features below. For concreteness, we will explain them in terms of an audio space system, like that described in [2], which applies a model of conversational floor participation in making participants in the same floor more intelligible to each other.

The first class of features consists of implicit or non-command interfaces, in which a system tracks "natural" human behaviors. Obviously, an audio conferencing system that tracks turn-taking (P5) [2] or a video conferencing system that tracks eye gaze [32] behavior to automate volume control is already an example of this. However, many refinements are possible, such as:

*Identifying potential SITs by detecting direct addressing* (P1). A turn beginning with something like "Carl, Carl" (Excerpt 1) is a potential SIT. Conventional speech recognition or wordspotting techniques can be applied to detect participants' use of proper nouns in the initial portion of a turn. Locating a potential SIT would not deterministically result in the model deciding that a schisming was in progress. Instead, the model would decrease its estimate of floor stability, making it more likely (during a limited time window) to decide that schisming had occurred.

*Identifying potential asides by detecting turns produced in overlap and with subdued delivery* (P3). Detecting overlap between participants and measuring speech energy are both relatively easy. However, this heuristic would be likely to be even more subject to "false alarms" than that of the previous example; again, this would only be an input used to adjust the model, not to make a deterministic decision to create a new floor.

*Identifying coordinated actions* (P6). Affiliating coordinated actions such as shared laughter involve correlations in time and speech energy. Unlike the previous two examples (which focused on schisming), detecting such correlations would result in the model increasing its estimate that the participants in question share a floor. As suggested in Section 4.2, this would be particularly useful in tracking secondary participants who say relatively little (such as A, S, T and W in Excerpt 4).

The second class of features consists of interfaces that actually provide different system features to different participants based on their inferred behavior. While the output of a multi-party interaction model might be some simple decisions about which person is in which floor, a probabilistic model will assign participants different likelihoods of floor affiliation or internal working categorizations (such as "possible secondary participant" – see above) based on their different conversational behaviors.

As a concrete example, a system could allow participants to "direct" their turns at certain groups of other participants without ever explicitly identifying them. Consider the fact that schisming based on toss-outs (P2), asides (P3) and retro-sequences (P4) often involves the recipiency of secondary participants rather than that of primary participants. Normally, any turns produced as part of a schisming would be heard at full volume (until the schisming succeeded) by all members of the floor. This does not help the schisming process and is disruptive to the primary participants. It would be interesting to support an abstraction such as "all of the secondary participants in my floor" as a group that can be easily addressed by other secondary participants. An aside (as discussed above) from a current secondary participant would result in that turn being heard at full volume only by this sub-group.

## 5.3 Modeling

Here, we discuss two general points relating to conversational modeling that have been highlighted by this research.

First, we (re-)emphasize the importance of multiple floors (Section 3.2) and multiple-party actions (P6). A system that assumes that conversational participation can be modeled as a single floor through observation of dyadic behaviors will not be useful in many real-world sociable interactions. For example, simply considering gap/overlap (P5) between pairs of speakers will fail when multiple turn-taking systems are operating.

Second, we note the need to model interactions with an eye to the fact that turns can have retroactive relevance.[2] Systems that analyze spoken language tend to look at it as a stream of features (machine learning) or phonetic/lexical/syntactic units (speech recognition). In interaction, the significance of a turn for the organization of a floor can be defined by other speakers' subsequent actions. This occurs through the simple uptake (or lack of uptake) of a turn, as with toss-outs (P2) and asides (P3). It can also be a radical transformation, as with retro-sequences (P4). This implies that fixed time-based analysis windows (as in [2,4,8,20]) should be extended in a way that ensures the continued consideration of relevant previous turns.

In the context of our own work, these points motivated a redesign of our adaptive audio space system. For the purpose of this discussion, we can characterize the original prototype [2] by three key attributes: it extracts features from a relatively short time window, considers only turn-taking features (P5), and identifies floors through analysis of (aggregated) pairwise measures of turn-taking. Informed by the work reported here, we are reengineering the system around a segment-based architecture, similar to that used in some speech recognition systems [12]. We expect this to improve on each of the three key attributes mentioned above; that is, it will provide a principled and more straightforward framework to consider data over longer time spans, to include features other than those based on turn-taking (such as potential schisming-related events (P1)-(P4)), and to analyze behaviors that span groups (such as coordinated actions (P6)).

## 6. CONCLUSION

Sociable interaction occupies an important place in daily experience, and as Simmel notes, its pleasures lie in the "ways in which groups form and split up and in which conversations, called forth by mere impulse and occasion, begin, deepen, loosen, and terminate," and in its role as "play [that] obeys the laws of its own form and whose charm is contained in itself" ([29], p.54). The work here contributes to our ability to conceptualize and model the processes of spontaneous, multi-party interaction. If "smart" applications of the kind we listed in the introduction (e.g., [4,5,8,20,21,27,30]) are to operate robustly in the real world, an environment in which highly spontaneous interaction occurs, their design should reflect these processes.

From a social science perspective, we have presented an applied conversation analytic study of sociable interaction within small groups. The contribution of this study is not in the "discovery" of "unknown" phenomena of conversation (in the sense that they do, after all, have plain-language names) but rather in their systematic collection and detailed characterization – in particular, the analysis of how each of the phenomena (excluding SITs [11]) shape and organize multi-party interaction. Given our interest in building systems that model the participation dynamics of such groups, we have steered the analysis in a direction that informs the design of such systems. As part of this, we have contributed *participation sequences* as an analytic construct.

From a technical perspective, the findings of this study have both specific and general implications. As discussed in Section 5.3, they have motivated a redesign of our own system. More generally, they have also resulted in a collection of user interface and user modeling ideas (Section 5.2) that can reused in other applications such as those described in the introduction.

Finally, we note that while the work here is presented in an applied context, we should not discount the value of basic research on multi-party interaction. To understand schismings and other phenomena, the sequential organization of their interactional environment must be analyzed in its own right.

---

[2] It is worth noting here that any interactive system that attempts to take action based on on-line conversation modeling will have some "lag" – that is, it is recognizing that floors are changing rather than predicting the changes, delaying actions by *at least* a partial turn. Our initial prototype studies suggest that this kind of lag is not *inherently* problematic for speakers [2] but this presumably has application-specific limits.

**Excerpt 1: schisming by *schism-inducing turn* (Lines 1-5); schisming by *toss-out* (Lines 6-13)**

```
1  N: ⇒ ca::rl, ca:rl, »carl carl carl,« so what'd dyou do this morning?,
2  C:                                                     this morning?
3  N:       [huh? (.) yeah,
4  C:       |           [this morn-? i- •h nohh, •h i: had a garage sale,
5  J:    did you [get to sleep [mo:re,
6  A:    i went home      [and my alarm clock was still on snooze,
7  M:                                         |            [∘heh heh heh∘
8  C:             [so i had to [wake  up      early,]        a garage [sale,
9  J:             [»you had to- (.) « you had to go where?]
10 C:    [∘i had to  wake up] early,∘
11 J:    o:::h, [i ha:te  tho:se, ] [i love having a ca:r] now, cuz i don't have to go to those,
12 A:       [it [wasn't- hhh it]hadn't] [turned back on yet, ]
13 M:             [tha:t's   cla:ssic,]
```

**Excerpt 2: schisming by *toss-out***

```
1  T:    my friend's getting ready to start his car=he's got a u:h sixty
2        eight nova with a: three eighty three (.) str[oker in [it?,]
3  H:                                         |         [tha [t's tight,
4  Z:                                                  [sounds   coo:::l,
5  S:    ni[:ce:,   ]
6  T:       [just hea]ders [right n]ow,]              [jim ]    ramirez,]
7  H:             w[ho?   ]  |[do you know [who?,]             |
8  Z:              [don't   know][ what it looks like, but it sounds] cool,
9  N: ⇒ •hh i need to s[ell    my    musta:ng,]
10 H:                  [did he just- did he j]ust buy it?,
11        (0.5)
12 J:    [d'n't  you] just get [your mustang?,
13 T:    [no,    ]             [oh they bought it a long time ago [but    they  (.)    tore] it apart
14 N:                                           [i've had for like a year,]
```

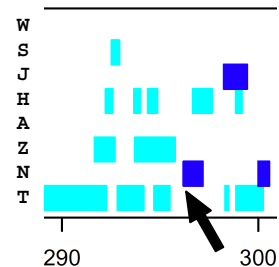

```
290                    300
```

**Excerpt 3: schisming by *aside***

```
1  H:    [I'll admit] that [like when I get] up in the chair
2  Z:              [ eh   huh ]
3        (0.3)
4  H:    [like]     [»or if someone ge[ts up in the
5  Z:                                 [   the: :
6  S: ⇒ [∘sur]prised the [Mexicans    can't       swim,
7  H:    chair to [bump me d]own like,«
8  Z:    chai: [  :   r,]
9  A:          [yeah      really,]
10 S:                  ya know,∘ ]
11 H:    lots of [black people there, they can't sw]im.
12 Z:    ∘ts[::: [  swam    over      h[ere,    ]
13 J:       [((laughter))          [that  i-] eh heh heh .HH @that's hella bad,@
```

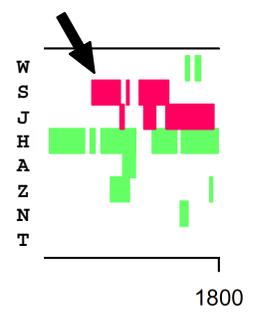

```
1800
```

**Excerpt 4:  schisming by *retro-sequence* (Lines 13-17); affiliation by *choral response* (Lines 1-12)**

```
1  N:    did the: lifeguards throw down?,
2        (0.2)
3  H:    no: like »some lady came up and was like what
4        are you guys gonna do and i was like« (.) watch.
5  J:    [((laughter))        ]
6  S:    [((laughter))        ]
7  H:    [((laughter))        ]
8  N:    [((laughter))        ]
9  W:    [((laughter))        ]
10 T:    [((laughter))        ]
11 A:    [((laughter))        ]
12 Z:    [@watch@ that's a great ][ one:, ]
13 N: ⇒                          [get my g]u:n?,
14 H:    i was like [o:h i'm calling in the police, but (.)  i'm basically gonna sit here and] watch,
15 Z: ⇒        [ho : [ : :h   yes : ]  :      that'd  be       awesome,]
16 W:              [stay         safe, ]
17 Z:    give all their lifeguard twelve gauge shot guns,
```

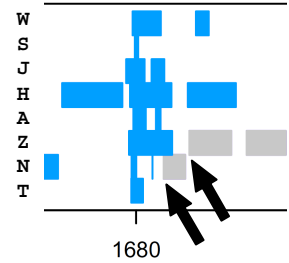

```
1680
```

**Table 3. Summary of transcription notation.**

| X: | X is speaking |
|---|---|
| Y: | *Y is speaking in a different floor* |

| My [talk]<br>  [your] talk | Alignment of overlapping speech or actions |
|---|---|
| (n)  (.) | *n* second pause; micropause |
| »quick«  °soft° | Rush-through; said softly |
| a: a | Elongated vowel; stressed speech |
| •hh  @haha@ | Audible inhalation; said laughingly |
| well, | Falling intonation |

# 7. ACKNOWLEDGEMENTS